\documentclass[pre,twocolumn,superscriptaddress,showpacs,preprintnumbers]{revtex4}

\newcommand{\bq}{\begin{equation}}
\newcommand{\ba}{\begin{eqnarray}}
\newcommand{\eq}{\end{equation}}
\newcommand{\ea}{\end{eqnarray}}

\newcommand {\tphi} {{\tilde \Phi}}

\newcommand{\calA}{\mathcal{A}}

\newcommand{\calD}{\mathcal{D}}

\newcommand{\Tr}[1]{\mathrm{Tr} [ \, #1 \, ]}

\usepackage{amsfonts}
\usepackage{amssymb}
\usepackage{amsmath}
\usepackage{feynmp}
\usepackage{color}
\usepackage{graphicx}
\begin{document}
\title{Auxiliary Field Loop expansion for the Effective Action for Stochastic Partial Differential equations II }
\author{Fred Cooper} \email{cooper@santafe.edu}
\affiliation{Department of Earth and Planetary Science, Harvard University,Cambridge, MA 02138}
\affiliation{The Santa Fe Institute, 1399 Hyde Park Road, Santa Fe, NM 87501, USA}

\date{\today}

\begin{abstract}
We extend our discussion of effective actions for stochastic partial differential equations to systems that give rise to a Martin-Siggia-Rose (MSR)  type of action.  This type of action naturally arises when one uses the many-body formalism of Doi and Peliti to describe  reaction-diffusion models  which undergo transitions into the absorbing state and which are described by a Master equation. These models include predator prey models, and directed percolation models as well as chemical kinetic models.  For classical dynamical systems with external noise it is always possible to construct an MSR action. 
 Using a path integral representation for the generator of the correlation functions, we show how, by introducing a  composite auxiliary field,  one can generate an auxiliary field loop expansion for the effective action for both types of systems. As  a specific example of the Doi-Peliti formalism we determine the  effective action for the  chemical reaction  annihilation and diffusion  process $A+A \rightarrow 0$.  For the external noise problem we  evaluate the effective action for the Cole-Hopf form of the  Kardar-Parisi Zhang (KPZ)  equation as well as for the Ginzburg Landau model of spin relaxation.   We determine for arbitrary spatial dimension $d$,  the renormalized  effective potential in leading order in the auxiliary field loop expansion (LOAF) and also determine the renormalization group equation for the running of the reaction rate (coupling constant)  for arbitrary $d$. We compare our results with known perturbative and non-perturbative results for the renormalization group equations.  
\end{abstract}
\maketitle

\section{Introduction}
The general theory of obtaining effective actions for stochastic partial differential equations has evolved in two directions.  One is based on the work of Onsager and Machlup  \cite{Onsager} \cite{Graham} \cite{Zinn-Justin}  where one directly uses the noise probability function to generate a path integral approach. The other is based on the response function formalism of Martin Siggia Rose (MSR)  \cite{msr} \cite{Peliti} \cite{janssen} \cite{Jouvet}, which gives an alternative path integral related to the Schwinger Keldysh closed time path formalism \cite{SK}.   If the stochastic dynamics is described by a Langevin equation whose noise is specified by a probability distribution function,
then one immediately obtains an MSR  Lagrangian by introducing a conjugate momentum field which is also a Lagrange multiplier field imposing the Langevin dynamics.
The MSR formalism also  naturally appears when one considers models of critical dynamics which arise from a Master equation.  These models are discussed in textbooks such as Kamenev \cite{Kamenev} as well as in the review article of Cardy \cite{Cardy}. 

  In the case that the Master equation represents the dynamics, then using the Doi-Peliti  \cite{Doi} \cite{Peliti} many body field theory approach, one is automatically led to a MSR type of Lagrangian.  This resulting dynamics for  the correlation functions  can  often be interpreted in terms of a Langevin dynamics but with complex noise (see for example the discussion of $A+A \rightarrow 0$ in the review of Cardy \cite{Cardy} ).  In our previous paper \cite{us} we showed how  to obtain an effective action of the Onsager Machlup type starting from the Langevin equation. In that paper,
 we discussed how to obtain the correlation functions of a system described by Langevin dynamics from the path integral representation for the generating functional. 
We also described how to introduce auxiliary fields into the problem so that we could evaluate the path integral  in a loop expansion in terms of the auxiliary field propagators. We then applied this method to the Kardar-Parisi-Zhang  (KPZ)  \cite{kpz}  equation where we derived the effective potential as a function of the dimension $d$.  Here  we continue our discussion of effective actions and discuss how to obtain the effective action using the MSR formalism.  This second formalism naturally arises when one is discussing birth and death processes coming from a Master equation and it is also the classical limit of the Schwinger-Keldysh formalism of quantum statistical dynamics \cite{Classical}. In most cases it is formally possible to go from one formulation to the other, modulo the issue of obtaining complex noise  \cite{Cardy} \cite{u1}.
 Once the MSR Lagrangian is obtained, it is again possible to introduce auxiliary fields which render the Lagrangian quadratic in the original fields (for a Classical dynamics example see \cite{pattern}). Having done that,  a loop expansion (in terms of loops containing the composite field propagator) of the generating functional for the correlation functions is again possible.  
 
 The MSR formalism lends itself to the possibility of introducing a 
{\it{different}} set of auxiliary fields than those suggested by the Onsager formalism. This comes about because there are new conjugate momentum fields  in the MSR version of the  Lagrangian.  We will display  such an example in terms of the Cole-Hopf transformed action for the KPZ equation,  which is related to the Action for the annihilation process $A+A \rightarrow 0$.  By introducing two  auxiliary fields in this reformulation of the KPZ action,  one of which is bilinear in the conjugate field, we are able to obtain a reasonable approximation to the exact renormalization group (RG) equation for the running of the dimensionless coupling constant. In our previous paper which used the Onsager-Machlup formulation \cite{us} we only were able to complete part of the renormalization program in leading order. 
 For pedagogic reasons,  we also determine the effective action for a system which relaxes at late times to thermal equilibrium and which is described by a Landau free energy which is quartic in the field $\phi$. This is the well known Ginzurg-Landau \cite{G-L}  model (see for example \cite{Cardy}).  In that system we derive the renormalized effective potential governing the dynamics of the relaxation process. We  also determine the  RG $\beta$ function as a function of $d$  and compare our results to those of perturbation theory.  It is important to point out that in the LOAF approximation the renormalization of the coupling constant is due to the point like ``scattering" being replaced by the exchange (or production and decay) of the composite auxiliary field.  Thus the renormalized coupling constant is simply related to the composite field propagator. 
This quantity is a renormalization group invariant, whose inverse can be easily determined from the second derivative of the effective action (or potential)  with respect to the auxiliary fields. 

In the recent literature  on stochastic partial differential equations most discussions of the effective action are based on a loop expansion in terms of the strength of the noise correlation function. This approximation is the familiar ``semi-classical" approximation described in standard field theory text books \cite{I-Z}.  In population biology, a recent discussion is found in \cite{Dodd} and in pair annihilation and Gribov processes a recent discussion is found in  \cite{Zorzano}. The usual loop expansion is related to a Gaussian truncation scheme  and is called the Bogoliubov approximation in the Bose-Einstein condensate literature  \cite{Andersen}. Its domain of validity is at weak coupling (and here also week noise strength).  By using an auxiliary field loop expansion which is closely related to the large-N expansion, we hope our results at lowest order will be useful at moderate values of the coupling (reaction rate). This extension of the domain of validity was observed when the  LOAF approximation was used to determine the phase structure of dilute Bose Gases \cite{BEC}. 

The paper is organized as follows.  In section II  we review  the path integral approach to stochastic partial differential equations of the reaction diffusion  type.
In section III we review the Doi-Peliti formalism for transforming Chemical Master equations to a many-body theory. 
 In section IV  we review the auxiliary field loop expansion method which has been used successfully in understanding BCS theory \cite{BCS}  and BEC  \cite{BEC}  theory as well as relativistic quantum field theories such as the scalar O(N) model  \cite{CJP} \cite{BCG} \cite{Nreview} . In section V  we obtain the effective action for annihilation chemical reaction  $A+A \rightarrow 0$. In section VI we obtain the effective potential for the annihilation process in the LOAF approximation.  In section VII we  relate the annihilation action to the Kardar-Parisi-Zhang  action in its Cole-Hopf form,  and derive an equation for the running of the dimensionless coupling constant.  In section VIII we obtain the effective action for the Ginzburg Landau model for  spin relaxation and also obtain the renormalization group equation for the running coupling constant.  Finally we summarize our results in section IX. 
 
\section{ Path integral formulation for Reaction diffusion equations with noise}
In this section we briefly review the path integral formulation for the correlation functions induced by external noise that has Gaussian correlations.  A generic system of coupled reaction diffusion equations with external noise can be  written in the schematic   form:
\bq
\frac{\partial {\Phi_i}}{\partial t} - \nu_i \nabla^2  \Phi_i - F_i[\Phi] - \eta_i \equiv D_i \Phi_i - F_i[\Phi] - \eta_i = 0. \label{phii}
\eq
We assume the noise is Gaussian in that the probability distribution function for noise can be described by 
\ba
&& P[\eta[x,t]] =  N \times \nonumber \\
&& \exp \left[ - \frac{1}{2} \int dx dydt dt'  \left[ \eta^i(x,t)  {G_\eta} ^{-1}_{ij} (x,t; y,t')   \eta^j(y, t') \right] \right] \nonumber \\
\ea
where $N$ is determined from
\bq
\int \prod_i {\cal D} \eta_i P[\eta]  = 1.
\eq
For this distribution
\bq
\langle \eta^i (x,t) \rangle  \equiv \int \calD \eta ~ P[\eta[x,t]] ~  \eta^i (x,t) =  0 ,
\eq
and the two point noise   correlation function (connected Green's function) is given by 
 \bq
\langle \eta ^i (x,t)   \eta ^j (y,t') \rangle_c   =   G_\eta  ^{ij}(xt; y t') .
\eq
We will assume that the strength of the noise correlation function is proportional to $\cal A$,  which is a parameter often used to control various approximation schemes such as loop expansions of the effective action in powers of $\calA$ \cite{Zorzano}.
In particular for our examples  we will choose 
\begin{equation}
G_\eta  ^{ij}(xt; y t') = {\cal A } \delta^d (x-x') \delta(t-t') .  
\end{equation}
  One assumes that for a particular configuration of the noise $\eta$, one can solve Eqs \ref{phii}
 for  $\Phi_i(x | \eta)$ and that there is a unique solution.  A strategy for doing this in the strong coupling domain is discussed in 
 \cite{bcf}.

The expectation values of the concentrations $ \Phi_i $ are obtained by performing the stochastic average over the noise

\bq \label{corr2}
\langle \Phi_i(x,t) \Phi_j(y,t') \rangle_{\eta}  = \int \calD  \eta P[\eta] \Phi_i(x,t | \eta) \Phi_j(y,t' | \eta ).
\eq

We are interested in getting a path integral representation for the correlation functions of Eq. (\ref{corr2}). 
Using the Fadeev-Popov trick we can, instead of explicitly solving for the $\Phi_i$ in terms of the noise $\eta$, enforce the fact that we are doing the integral over fields that obey the noisy reaction diffusion equation.  That is we insert the functional delta function into the path integral over $\eta$ using the identity:
\ba \label{delta}
&&1= \int \calD \Phi_i  \delta [\Phi_i - \Phi_i (x| \eta)] \nonumber \\
&&=\int \calD \Phi_i  \delta [\partial {\Phi_i}{\partial t} - \nu_i \nabla^2  \Phi_i - F_i[\Phi] - \eta_i]  | {\rm det}  S^{-1}[\Phi] | . \nonumber \\
\ea
Here 
\bq 
S^{-1}_{ij}  =  \left( D_i \delta_{ij}  -  \frac {\partial F_i[\Phi] }{\partial \phi_j } \right) \delta(x-y) \delta(t-t') .
\eq
The determinant can be replaced by a path integral over fermionic fields or ignored if we use an appropriate choice of the
lattice version of the time derivative (forward derivative) which is the Ito regularization  \cite{Cardy} \cite{Kamenev}. 
At this point there are two approaches to obtaining a field theory description.  One can directly integrate over the noise to obtain
the approach of Onsager and Machlup \cite{Onsager}  \cite{Graham} \cite{Zinn-Justin} . This is the approach we took in our previous paper \cite{us}.

 The approach we will take  in this paper is to instead  introduce a functional representation for the delta function to obtain a Martin-Siggia-Rose  action \cite{msr}  \cite{janssen} \cite{Jouvet} \cite{Kamenev} which is closely related to the Schwinger-Keldysh formalism \cite{SK}.

The generating functional  $Z[j]$ for the correlation functions is given by
\bq
Z[j] =\int \calD \eta  P[\eta] \exp[\int  dx J_i(x) \Phi_i (x | \eta) ]  .
\eq

We represent the functional delta function in Eq. \eqref{delta} by a Laplace transform over a Lagrange multiplier field $\tphi$. 
\ba
 &&\delta [\partial {\Phi_i}{\partial t} - \nu_i \nabla^2  \Phi_i - F_i[\Phi] - \eta_i]  =  \nonumber \\
&& \int \calD \tphi \exp[ - \int dx \tphi_i(x) \left( \partial {\Phi_i}{\partial t} - \nu_i \nabla^2  \Phi_i - F_i[\Phi] - \eta_i\right).  \label{laplace} \nonumber \\
 \ea

We next insert Eq. \eqref{delta} into Eq. \eqref{zj}. 
Adding sources $J_i$ and ${\tilde J}_i$  for the fields $\Phi_i$ and $\tphi_i$ we find that the generating functional for the connected Green's function is given by
\ba
Z[J,{\tilde J}] &&= \int \calD \eta \calD  \Phi \calD  \tphi  \exp  \left[- S_1[\Phi, \tphi, \eta]  \right. \nonumber \\
&& \left. +  \int dx  ( J_i \Phi_i + {\tilde J}_i \tphi_i) \right] . \label{zj}
\ea
Here
\ba
S_1 &&=  \int dx \tphi_i(x) \left( \partial {\Phi_i}{\partial t} - \nu_i \nabla^2  \Phi_i - F_i[\Phi] - \eta_i  \right) \nonumber \\
&&
 - \frac{1}{2} \int dx dy \left[ \eta^i(x)   (G_\eta)^{-1}  _{ij} (x,y)  \eta^j (y) \right] .
\ea

We can now perform the Gaussian integral over the noise to obtain 
\ba
&&Z[J,{\tilde J}] =  \nonumber \\
&& \int \calD  \Phi \calD  \tphi  \exp  \left[- S_2[\Phi, \tphi] +  \int dx  ( J_i \Phi_i + {\tilde J}_i \tphi_i) \right]. \nonumber \\
\ea
\ba
S_2&&= \int dx \tphi_i(x) \left(\partial {\Phi_i}{\partial t} - \nu_i \nabla^2  \Phi_i - F_i[\Phi]  \right)  \nonumber \\
&&- \frac{1}{2} \int dx dy  \tphi_i(x)  (G_\eta) _{ij} (x,y) \tphi_j(y) .   \label{gen}
\ea

At this point we want to point out two facts.  If we now perform the path integral over the field $ \tphi_j(y)$, then we recover the Onsager-Machlup form of the generating functional. 
Another way to obtain the Onsager-Machlup action is to consider the field $ \tphi_j(y)$ as a Lagrange multiplier field to be determined by the relationship
\bq
\frac{\partial S_2}{\partial \tphi_j(y)} =0.
\eq
One then finds 
$
S_2  [ \Phi,  \tphi [\Phi] ]
$ is the Onsager-Machlup form of the action which then can be used to determine the $\Phi$ correlation functions.

\section{Master Equation and Many Body Formalism} 
The master equation arises in many situations that can be described in terms of birth and death processes.  Here we will assume that we are describing chemical reactions, since we are going to consider the annihilation process $ A+ A \rightarrow 0$ which is often discussed in the literature \cite{Lee},  \cite{THVL_2005} \cite{Cardy}. 
In order to develop the master equation formalism for a system of chemical reactions, we first divide the space in which the reactions take place into a $d-$dimensional hyper-cubic lattice of cells and assume that we can treat each cell as a coherent entity. We assume the interactions occur locally at a single cell site and that there is also diffusion modeled as hopping between nearest neighbors. Assuming that the underlying processes are Markovian and describable by a probability distribution function $P(\mathbf{n} ,t)$ which gives the  probability to find the particle configuration $\mathbf{(n)}$ at time $t$.
Here  $\mathbf {n_i(t)} =( \{n_i(t) \}) $ is  a vector composition variable where $n_i$  represents the number of molecules of a species at site $i$. Let us look at a simple example for the chemical reaction $A+ A \rightarrow 0$. 
The master equation for annihilation at a single site with reaction rate $\lambda$  is given by  \cite{Cardy} 
\bq
\frac{dP(n)}{dt} = \lambda (n+2) (n+1)P(n+2)  - \lambda n( n-1) P(n).
\eq
Diffusion is modeled by simple hopping from nearest neighbor sites. For example if we consider two sites $ (1,2)$ and hopping  $1 \rightarrow 2$ at a rate $D$
\bq
\frac{dP(n_1,n_2)}{dt}  = D(n_1+1) P(n_1+1, n_2 -1) - Dn_1 P(n_1,n_2).
\eq
We then include hopping also from $2 \rightarrow 1$ at the same rate. 

The master equation lends itself  to a many body description \cite{Doi}, accomplished by the  introduction of an occupation number algebra with annihilation/creation operators
$\hat a_i, \hat a_i^\dag$  for $A$ at each site $i$. These operators obey the Bosonic commutation relations
\ba
\left[\hat a_i, \hat a^\dag _j \right] &=& \delta_{ij}\nonumber \\
\left[\hat a_i,\hat a _j \right] &=& 0, \quad~\left[\hat a_i^\dag,\hat a^\dag _j \right] =0,
\label{commute}
\ea
and define the occupation number operators $\hat n_{i} = \hat a_i ^{\dag} \hat a_i$ satisfying the following eigenvalue equations:
\begin{equation}
\hat n_{i} | n _{i}\rangle = n_{i} | n _{i,}\rangle.  
\label{eigen}
\end{equation}
We next construct the state vector
\bq  \label{wavefunction}
|\Psi(t) \rangle =\sum_{n_1,n_2}   \hat a_1^{\dag  n}  \hat a_2^{\dag  n}  P(n_1,n_2 ,t)  | 0 \rangle ,
\eq
which upon differentiating with respect to time $t$, can be written in the suggestive form
\bq  \label{Schrodeq}
-\frac{\partial |\Psi(t) \rangle }{ \partial t} = H [ {\hat a_i ^{\dag}},{\hat a_j}] \Psi(t) | \rangle,
\eq
resembling the Schr\"odinger equation. Finally, taking  the time derivative of Eq. (\ref{wavefunction}) and comparing terms with the Hamiltonian in~\eqref{Schrodeq} we make the identification
\ba  \label{Hsym}
H &=& D \sum_{\langle i,j \rangle}  (\hat a_i^\dag - \hat a_j^\dag)( \hat a_i- \hat a_j)  -{\lambda} \sum_i  \bigl[ \hat a_i^{2} - \hat a_i^{\dag2} \hat a_i^{2}  \bigr]  . \nonumber \\
\ea

Having defined the space, the appropriate wave function and the  Hamiltonian, we next seek to evaluate the operator $ e^{- \tilde H t} $ using the path integral formulation. 
We follow the standard procedure for obtaining the coherent state path integral~\cite{NO98, THVL_2005, Tauber_2007,review} . Letting the coherent state $\phi(x)$  represent the eigenvalue of  $a_i$  and also let $\phi^\star(x) $  represent  $a_i^\dag$  we obtain
\bq
e^{-\tilde Ht} = \int \calD \phi\calD\phi^\star e^{- S[\phi^\star \phi]}, \label{pathint}
\eq
where the action $S$  is given by
\ba
S = && \int dx \int_0^\tau dt \bigl[ \phi^\star \partial_t \phi+  \nu \nabla \phi^\star \nabla \phi   -  \lambda \phi^2 (1-(\phi^\star)^2)   \bigr] . \nonumber \\
 \label{action}
\ea

This can also be represented by a Langevin equation.  To see this we make the Doi shift
\bq
\phi^\star \rightarrow 1+ \phi^\star ,
\eq
which leave the kinetic terms unchanged. The Doi-shifted action is now 

\ba
S = && \int dx \int_0^\tau dt \bigl[ \phi^\star \partial_t \phi+  \nu \nabla \phi^\star \nabla \phi   + 2  \lambda \phi^2 + \lambda \phi^2 (\phi^\star)^2)   \bigr] . \nonumber \\
 \label{actiondoi}
\ea
The last term in the action can be obtain from the identity
\ba
&& \exp[ -\int dx dt \lambda \phi^2 (\phi^\star)^2) ] =  \nonumber \\
&&\int \calD \eta P[\eta] \exp \int dt dx \{i   \phi^ \star  \sqrt{2 \lambda} \phi\eta \}  , \nonumber \\
\ea
where the noise is Gaussian, with 
\bq
\langle \eta(x,t) \eta(x',t') \rangle = \delta^d (x-x') \delta(t-t').
\eq
The Doi-Shifted  path integral can be obtained using the previous approach starting with the Langevin equation
\ba
\partial_t \phi && = \nu \nabla^2 \phi - 2 \lambda \phi^2 + i  \sqrt{2 \lambda} \phi   \eta . \nonumber \\
\ea
This shows that  the noise source is multiplicative and pure imaginary.  If we generalize the chemical kinetics so that there are three molecules involved in the chemical reaction, such as  in the annihilation reaction $3 A  \rightarrow 0$, one needs to introduce auxiliary fields in order to obtain a Langevin description \cite{u1} . 

\section{Auxiliary field loop expansion}
Now let us imagine that by means of introducing as set of  auxiliary fields $\sigma_k$ (which also includes $\tilde \sigma_k$ )   the action is now rendered quadratic in the original fields $\phi_i $  (which symbolically stands for the combination $\phi_i, \tilde \phi_i$).  This can be done by  a Hubbard-Stratonovich transformation \cite{Hubbard,BCG,CJP},  or using a functional delta function using yet another auxiliary Lagrange multiplier field \cite{Nreview}.   After doing this,  the ``bare"  inverse propagator for the $\phi$ fields can be symbolically written as 
$G_\phi  ^{-1}[\sigma] $ 
and the full action can be written as 
\ba
S [\phi, \sigma]&& = \int  dx dy  dt dt' \left[  \phi_i[x,t, ]{ G^{-1}_\phi}  ^{ij}  (x,y, t, t'; \sigma)  \phi_j [y,t']   \right. \nonumber \\
&& \left. + S_2 [\sigma_i]  \right] -  \int dx dt [j_i \phi_i+ J_k \sigma_k ],
\ea
where now $S_2[\sigma]$ is only a function of the auxiliary  fields $\sigma_i$. 

After introducing the auxiliary fields, the generating function for the noise induced interactions (see Eq. \eqref{zj})  is  now of the form:
\bq
Z[{J,j} ] = \int \calD \sigma  \calD \phi  \exp \left[ - { S}[\phi, \sigma  ] \right] .  \label{gen2}
\eq

One can now perform the Gaussian Path  integral over $\phi$ to obtain an equivalent  action ${\hat S}$  which just depends on the field $\sigma$ and the external 
sources $J, j$. 
Performing the Gaussian integral, we are left with the  expression for the generating functional:
\ba
Z[{J,j}]&& = \int \calD \sigma  \exp [ - {\hat S}[\sigma,j, J]  ] ,  \label{seff1}
\ea
where
\ba
{\hat S}&& =  \int dx  S_2[\sigma] -  \int dx dy \frac{1}{2}  j(x)   G(x,y; \sigma) j(y)  \nonumber \\
&&- \int dx J(x) \sigma(x)  +  \frac{1}{2} ~ {\rm Tr} ~\ln G^{-1} (\sigma) ,   \nonumber \\  \label{seff2a}
\ea
and we have added a source term $J$  for  the auxiliary field $ \sigma$.  $W[J, j] = \ln Z[j, J] ] $ is the generator of the connected correlation functions.

We next introduce a small parameter $\epsilon$  into the theory  via the substitution
${\hat S} \rightarrow {\hat S}/\epsilon$.  For small $\epsilon$ evaluation of the integral by steepest descent (or Laplace's method)  is justified. $\epsilon$ is similar to $\hbar$ in that it counts loops but now the loops are in the propagators for the auxiliary fields $\sigma$. 
The auxiliary field loop expansion is obtained by first  expanding  around the stationary phase point $\sigma_0$, and using the Gaussian term for the
measure of the remaining integrals expanded as a power series in $\epsilon$ \cite{BCG}.
The details of obtaining the loop expansion are in our previous paper \cite{us} as well as in \cite{BCG} \cite{BEC2} and we will just summarize the results here.  We evaluate the path integral by steepest descent and keep terms to order $\epsilon$.   Next we expand the effective action about the  point $\sigma_0^i$, and evaluate the path integral by the method of steepest descent (or Laplace's method).  
Keeping terms up to quadratic in the expansion about the stationary phase point $\sigma_0$ and performing the Gaussian integral in $\sigma$   we obtain for the $W[J,j] $

\bq
-W[J,j ] =  {\hat S}[\, \sigma_0,J \,] + \frac{\epsilon}{2} \Tr {\ln D_\sigma ^{-1} [\sigma_0]} ,
\eq
where 
\bq \label{Dinv}
 D_{ij}^{-1} [\sigma_0](x,x') = \frac{\delta^2 {\hat S}[\, \sigma,J \,]}
        {\delta \sigma^i(x) \, \delta \sigma^j(x')} \Big |_{\sigma_0} 
\eq
is the inverse propagator for the composite $\sigma$ field in the leading order. 
Higher terms in the loop expansion in the  $\sigma$ propagator are obtained by treating the higher terms in the derivative  expansion of $ {\hat S}[\sigma]$ perturbatively with respect to the Gaussian measure.  This is discussed in detail in \cite{BCG}.  

  We then make a Legendre transformation from the external sources $j, J$ to the expectation value of the fields $ \phi, \sigma$ via

 \bq
\Gamma [\phi, \sigma] =  -\ln Z[J,j ] + \int  dx ( j  \phi+J \sigma ) ,
\eq
where here
 
\bq
\phi = \langle \phi \rangle   =  \frac{ \delta  \ln Z[J] } {\delta j} ; ~~ \frac{ \delta \Gamma [ \phi, \sigma] } {\delta \phi} = j.
\eq

\bq
\sigma = < \sigma>  =  \frac{ \delta  \ln Z[J,j ] } {\delta J} ; ~~ \frac{ \delta \Gamma [ \phi,  \sigma] }{\delta \sigma} = J.
\eq
$\Gamma [ \phi, \sigma ]$,  is the generator of the one-particle irreducible graphs,  and is the Legendre Transform of  $Z[j, J]$.

 The effective potential,
which generalizes the idea of a potential in classical mechanics,  is the value of $\Gamma$ for constant fields divided by the space-time  volume $\Omega$. 
Again if we keep the stationary phase point plus Gaussian fluctuations we obtain schematically for $\Gamma$ up to order $\epsilon$
\ba  \label{effact1}
\Gamma [\phi, \sigma] && = \frac{1}{2}  \int  dx dy  dt dt' \left[  \phi[x,t, ] G_\phi^{-1}(x,y, t, t'; \sigma)  \phi[y,t']  \right. \nonumber \\
&& \left. + S_2 [\sigma]  \right] 
 + \frac{1}{2} \Tr {\ln G_\phi^{-1} [\sigma ]} \nonumber \\
&&  +  \frac{\epsilon}{2} \Tr {\ln D_\sigma^{-1} [\sigma, \phi]}.
\ea
The LOAF approximation consists of just keeping the stationary phase part of the effective action (i.e. $\epsilon \rightarrow  0$) . 

\section{Effective Action for  {$ \bf{A+A \rightarrow  0}$}}

The effective action for the annihilation process $A+A \rightarrow 0$   has been derived in a one loop approximation using the expansion parameter $\calA$   by Hochberg and Zorzano \cite{Zorzano} .  In their calculation $\cal A$ counts loops in an expansion around the semiclassical action, in distinction to $\epsilon$ which counts loop corrections to  the self consistently determined mean field action. The LOAF approximation which we will derive here is non-perturbative in $\cal A$ but instead perturbative in the artificial  auxiliary field propagator loop counting parameter $\epsilon$.   Hochberg and Zorzano were able to explicitly calculate the effective potential only in the critical dimension $d=2$.  In that dimension our results for the renormalization group equation  (RGE) re-expanded to leading order in $\calA$ agrees with theirs.  However,  because of the simpler way terms are grouped in our approach we are able to determine the renormalized effective potential at arbitrary $d$ .  We therefor are able to  obtain 
an RGE equation for the running of the reaction rate which quantitatively agrees with the exact answer at all $d$.  We also obtain an approximate determination of the infrared stable fixed point when $d<2$. 

  The starting point for our calculation is the path integral representation for the generating functional for the correlation functions
\bq
Z[j,J]  = \int \calD \phi\calD\phi^\star e^{- S[\phi^\star \phi] + \int dx ( j^\star \phi + \phi^\star j)  }, \label{pathint1}
\eq
where the action $S$  is given by
\ba
S = && \int dx \int_0^\tau dt \bigl[ \phi^\star \partial_t \phi+  \nu \nabla \phi^\star \nabla \phi   -  \lambda \phi^2 (1-(\phi^\star)^2)   \bigr] . \nonumber \\
 \label{action1}
\ea
We now introduce the composite fields $\sigma$ and $\sigma^\star$ via a Hubbard-Stratonovich transformation.  That is we add to the above Lagrangian the action
\bq
S_{HS}  = \int dx dt \left[  - \frac{1}{\lambda } (\sigma^\star - \lambda \phi^\star \phi^ \star)(\sigma - \lambda \phi \phi) \right].
\eq
Adding this to Eq.  \eqref{action1}  we get an  action which is now quadratic in $\phi, \phi^\star$.
\ba
S_{2}[\phi,\sigma] && = \int dx \int_0^\tau dt \left[ \phi^\star \partial_t \phi+  \nu \nabla \phi^\star \nabla \phi  +    (\sigma^\star - 1) \phi^2 \right. \nonumber \\
&& \left.  +    \sigma (\phi^\star)^2)     - \frac{1}{\lambda} \sigma^\star \sigma \right] .
 \label{action2}
\ea
The ``bare''  inverse propagator for the matrix $\phi$ field can be written 
\ba
&&G^{-1}(x-x',t-t'; \sigma, \sigma^\star) =   \nonumber \\
&&\left(  \begin{array}{cc}
   2  \sigma   & - \partial_t - \nu \nabla^2 \\ 
  \partial_t - \nu \nabla^2     & 2 (\sigma^\star -1)    \\ 
  \end{array}
\right)  \delta^d (x-x') \delta (t-t') . \nonumber \\
\ea

Performing the integration over the fields $\phi, \phi^\star$ and adding sources for the auxiliary fields we obtain
\ba
&&Z[j,j^\star, J, J^\star]=  \nonumber \\
&&\int \calD \sigma \calD \sigma^\star \exp \left[- \left( \tilde S [ \sigma,\sigma^\star, j, j^\star] - J^\star \sigma - \sigma^\star J \right) \right] , \nonumber \\
\ea
where schematically: 
\bq
\tilde S = - \int \left( j \cdot G \cdot j + \lambda \sigma^\star \sigma \right)  + \frac{1}{2}  \rm {Tr}   \ln { G^{-1}(\sigma,\sigma^\star) }. 
\eq
 Performing the integration over the auxiliary fields by steepest descent and keeping the saddle point contribution, we can then Legendre transform the result for $\ln Z$ and we obtain in leading order in the auxiliary field loop expansion:
\ba
\Gamma [ \phi, \phi^\star, \sigma, \sigma^\star]&& = - \ln Z[J, J^\star, j, j^\star] \nonumber \\
&& + \int dx [ j^\star \phi +  j \phi^\star + \sigma^\star J + \sigma J^\star] \nonumber \\
 &&= S_2[\phi, \sigma] + \frac{1}{2} \rm {Tr}   \ln { G^{-1}(\sigma,\sigma^\star) }. 
\ea
Here $\Gamma$ is the generating functional of the one particle irreducible graphs.  Now that we have $\Gamma$ we can obtain the equations of motion  for the time evolution of the expectation value of the field from 
\bq
\frac{ \delta \Gamma} { \delta \phi^\star} =0.
\eq
The two particle correlation functions are obtained from the inverse of the Matrix of second derivatives of the effective action with respect to the fields.  In this
way one can look at the noise averaged dynamics of the correlation functions.  In this paper we will concentrate on the phase portrait obtained from the effective potential and the renormalization group flow of the running coupling constant. 

\section{Effective Potential}

Restricting ourselves to constant fields and defining the effective potential as 
\bq
V_{eff} = \frac{ \Gamma [\phi,\phi^\star, \sigma, \sigma^\star]} {\Omega} ,
\eq
where $\Omega$ is the space time volume we obtain 
\bq
V_{eff}  =    (\sigma^\star - 1) \phi^2 +  \sigma (\phi^\star)^2)     - \frac{1}{\lambda} \sigma^\star \sigma
	  +\frac{1}{2} \rm {Tr}   \ln { G^{-1}(\sigma,\sigma^\star) } ,
\eq
which contains the static part of the classical action plus the saddle point contribution to the fluctuations. For static fields we can Fourier transform the propagator and evlauate the fluctuations by calculating the log of the determinant of $G^{-1}$ in momentum space.
In momentum space we have 
\bq
G^{-1}(\omega, q, \sigma, \sigma^\star) =  \left(  \begin{array}{cc}
   2 \sigma   &~~ - i \omega + \nu k^2 \\ 
i \omega + \nu k^2      &~~ 2 (\sigma^\star -1)    \\ 
  \end{array}
\right) .
\eq

Evaluating the determinant we obtain
\ba
V_{eff} && =    (\sigma^\star - 1) \phi^2 +  \sigma (\phi^\star)^2)     - \frac{1}{\lambda}  \sigma^\star \sigma \nonumber \\
 &&+ \frac{1}{2} \int \frac{d^d k}{(2 \pi)^d} \int_{-\infty}^{\infty} \frac{d \omega}{2 \pi} \ln\left[\omega^2 +\nu^2 k^4 + 4 \sigma(1-\sigma^\star)  \right] .\nonumber \\
\ea
Performing the integration over $\omega$ we obtain
\ba
V_{eff} && =   (\sigma^\star - 1) \phi^2 +  \sigma (\phi^\star)^2)     - \frac{1}{\lambda}  \sigma^\star \sigma \nonumber \\
&& + \frac{1}{2} \int \frac{d^d k}{(2 \pi)^d} \sqrt{\nu^2 k^4 + 4  \sigma(1-\sigma^\star)} .
\ea
This expression needs to be evaluated at the solution of the gap equations 
\bq
\frac {\partial V}{\partial \sigma} = \frac {\partial V}{\partial \sigma^\star} = 0.
\eq
We have:
\bq
\frac {\partial V}{\partial \sigma^\star}  =  \phi^2 - \frac{1}{\lambda}  \sigma -  \sigma \int \frac{d^d k}{(2 \pi)^d} \frac{1} {(\nu^2 k^4 + 4  \sigma(1-\sigma^\star))^{1/2}}
\eq

\ba
\frac {\partial V}{\partial \sigma} && =  (\phi^\star)^2 - \frac{1}{\lambda} \sigma^\star  \nonumber \\
&& +( 1-\sigma^\star)  \int \frac{d^d k}{(2 \pi)^d} \frac{1} {(\nu^2 k^4 + 4 \ \sigma(1-\sigma^\star))^{1/2}}. \nonumber \\
\ea

The second derivative  of the potential is related to the  inverse of the  correlation function $D^{-1}_{\sigma \sigma^\star}$  of the $\sigma$ field at zero momentum. We have
\ba
-D^{-1}_{\sigma \sigma^\star} &&=\frac {\partial^2 V}{\partial \sigma \partial \sigma^\star }   \nonumber \\
&& = - \frac{1}{\lambda}  -\frac{1}{\nu} \left( \Sigma_1[m^2]  -  \Sigma_2[m^2] \right).
\ea
Here 
\ba
m^4 &&= \frac{4 \sigma(1-\sigma^\star)} {\nu^2}; \nonumber \\
 \Sigma_1[m^2]&&= \int \frac{d^d k}{(2 \pi)^d} \frac{1} {(k^4 +  m^4)^{1/2}} , \nonumber \\
 \Sigma_2[m^2]&& = \frac{m^4}{2}  \int \frac{d^d k}{(2 \pi)^d} \frac{1} {(k^4 + m^4)^{3/2}} .
\ea
We notice that $\Sigma_1$ has ultraviolet divergences in dimensions two and higher.  Let us introduce a mass scale where we define the renormalized reaction rate.  The running coupling constant in the theory rewritten in terms of $\sigma$ and $\sigma^\star$ is 
\bq
\lambda_r[q,\omega,\mu^2] =  D_{\sigma \sigma^\star}[q, \omega, \mu^2] .
\eq
This is a renormalization invariant. Here $\mu^2$ is the reference mass squared at which we choose $m^2$. Specifically we have the equation
\bq
\frac{1}{\lambda_r[\mu^2] } =   D^{-1} _{\sigma \sigma^\star}[0, 0, \mu^2]    \equiv  \frac{1}{\lambda} +  \frac{1}{\nu} (\Sigma_1[\mu^2]-\Sigma_2[\mu^2]) .
\eq
We have that 
\ba
\Sigma_1[m^2] && =  m^{d-2} \Omega_d  \frac{\Gamma \left(\frac{1}{2}-\frac{d}{4}\right) \Gamma 
   \left(\frac{d}{4}\right)}{4 (2 \pi)^d \sqrt{\pi }} \nonumber \\
&& =\frac{1}{2-d}  m^{d-2} \Omega_d  \frac{\Gamma \left(\frac{3}{2}-\frac{d}{4}\right) \Gamma 
   \left(\frac{d}{4}\right)}{ (2 \pi)^d \sqrt{\pi }}.
   \ea
\bq
\Sigma_2 [m^2]= m^{d-2}  \Omega_d \frac{\Gamma \left(\frac{3}{2}-\frac{d}{4}\right) \Gamma \left(\frac{d}{4}\right)}{4 (2 \pi)^d
   \sqrt{\pi }}.
   \eq
 The angular integration factor in $d$ dimensions is
 \bq
 \Omega_d = \frac{ 2 (\pi)^{d/2}  }{\Gamma(d/2)}.
 \eq
 The scattering amplitude  (renormalized decay rate) defined at scale $\mu$ is given by
 \bq
 \lambda_r[\mu] = Z_g[\mu] \lambda;, Z_g^{-1} =  1+ \frac{\lambda}{\nu} \Sigma[\mu^2]. \label{scat}
 \eq
 Here 
 \bq
 \Sigma[m^2 ]  = \Sigma_1[m^2]- \Sigma_2 [m^2] =   m^{d-2} \frac{1}{2-d}  H[d] ,
 \eq
 where
 \bq
H[d] =\frac{2^{-d-1} (d+2) \pi ^{-\frac{d}{2}-\frac{1}{2}} \Gamma
   \left(\frac{3}{2}-\frac{d}{4}\right) \Gamma \left(\frac{d}{4}\right)}{\Gamma
   \left(\frac{d}{2}\right)} .
    \eq
 Therefore we get the renormalization group equation for $\lambda_r$ : 
 \bq
 \mu  \frac { d \lambda_r}{d \mu} = \frac{\lambda_r^2}{\nu} H[d] .
 \eq

Another way of looking at Eq.  \eqref{scat} is to compare $\lambda_r$ at two different mass scales. Then one obtains
\bq
\frac{1}{\lambda_r[\mu^2]} - \frac{1}{\lambda_r[\mu_0^2]}  = \frac{H[d]} {\nu (2-d) } \left[ \mu^{d-2} - \mu_0^{d-2} \right],
\eq
or equivalently
\bq
\lambda_r[\mu^2] = \frac{\lambda_r[\mu_0^2] }{ 1+ \frac{H[d]} {\nu (2-d) } \left[ \mu^{d-2} - \mu_0^{d-2} \right]}. \label{run5}
\eq
 We can form the dimensionless bare decay rate
 \bq
 g_0 = \frac{\lambda \mu^{d-2}}{\nu} ,
\eq
in terms of which the dimensionless renormalized reaction rate is given by 
\bq
g_r[\mu^2] = \frac{g_0}{1+g_0  \frac{H[d]}{(2-d)}} .
\eq

The beta function is
\bq
\beta_g = \mu \frac{\partial g_r}{\partial \mu} = (d-2) g_r + g_r^2 H[d] .
\eq
This is just the differential form of Eq. \eqref{run5}.
The exact $\beta$ function is calculated by summing all the perturbative one loop graphs (see for example \cite{Lee}).  This leads to the exact answers:
\bq
g_r[\mu^2] = \frac{g_0}{1+g_0  \frac{B[d]}{(2-d)}} ,
\eq
\bq
\beta_g = \mu \frac{\partial g_r}{\partial \mu} = (d-2) g_r + g_r^2   B[d],
\eq
where 
\bq
 B[d] = 2! 2^{-d/2} (4 \pi) ^{-d/2} \Gamma(2-d/2).
 \eq
 
\begin{figure}[t!]
    \centering
   \includegraphics[width=0.8\columnwidth]{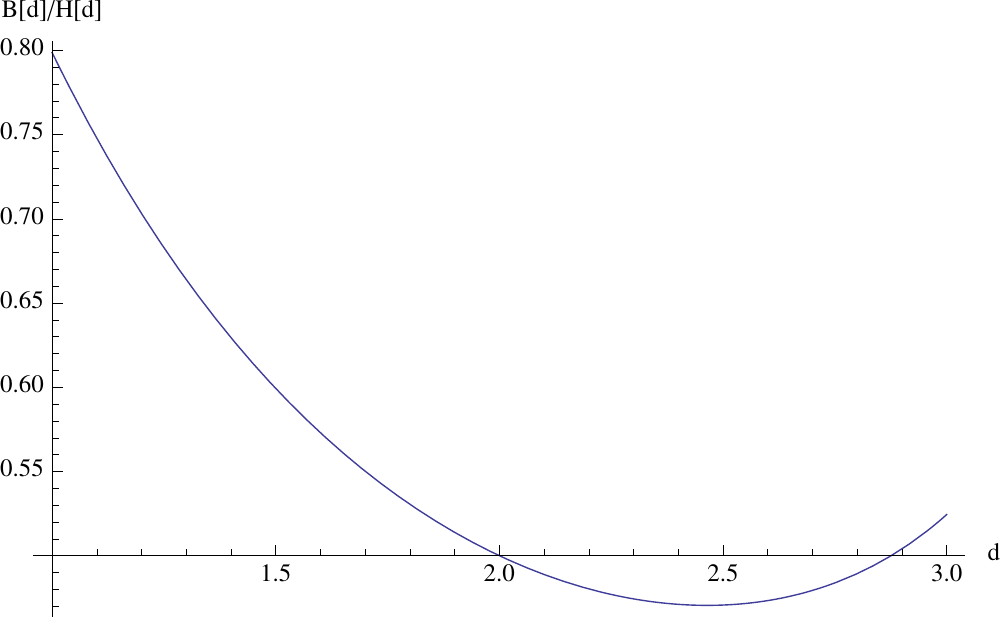}
    \caption{
   The ratio $B[d]/H[d]$ as a function of $d$.}
   \label{fig:fig1}
    \end{figure}

 The ratio of $B[d]$ to $H[d]$ is slowly varying from $d=1$ to $d=3$, (see Fig. \ref{fig:fig1}) showing that we get a reasonable answer in our LOAF approximation to the running of the coupling constant. 

Our result (as well as exact result)  shows that for $d<2$ there is a stable infrared fixed point
\bq
g^\star = \frac{2-d} {H[d]} .
\eq
In terms of  the second derivative of $V_{eff}$ at the reference mass $\mu$
\bq
 \frac {\partial^2 V_{eff}}{\partial \sigma \partial \sigma^\star }|_\mu   =-\frac{1}{\lambda_r[\mu^2]}, 
\eq
we can write the second derivative  as 
\bq 
- \frac {\partial^2 V_{eff}}{\partial \sigma \partial \sigma^\star } = \frac{1}{\lambda_r[\mu^2]} + \frac{H[d]}{\nu (2-d)}[(m^4)^{(d-2)/4}-(\mu^4)^{(d-2)/4}] . \label{second}
\eq

To obtain the renormalized effective potential we just need to integrate with respect to $\sigma$  and $\sigma^\star$
being mindful of the constants of integration.
We find
\ba
V_{eff} && =   (\sigma^\star - 1) \phi^2 +   \sigma (\phi^\star)^2)     - \frac{1}{\lambda_r[\mu^2]} \sigma^\star \sigma \nonumber \\
&&+I[d]   \sigma (\sigma^\star -1) \left(\mu ^{d-2}-\frac{16} {(d+2)^2} m^{d-2} \right) ,
 \ea
  where  $I[d] = \frac{H[d]}{\nu (d-2)}$, and $m^4 = \sigma (1-\sigma^\star)/\nu^2$.

This has to be evaluated at the solution to the two gap equations:
\bq
\frac{\partial V_{eff} } {\partial \sigma} = 0,  ~~\frac{\partial V_{eff} } {\partial \sigma^\star}=0.
\eq

In the critical dimension $d=2$ we can expand $\Sigma$ in  powers of $\epsilon = 2-d$ to obtain
\ba
\Sigma[ \mu] &&= \frac{-4 \log (\mu )-3 \gamma -1+2 \log ( 4 \pi )-\psi
   ^{(0)}\left(\frac{1}{2}\right)}{8 {\pi }} \nonumber \\
  && +\frac{1}{2 {\pi } \epsilon } + O( \epsilon^2). \label{exp}
   \ea
Using Eq. \eqref{exp}    we obtain from Eq. \eqref{second} in the limit that $d \rightarrow 2$
\bq 
- \frac {\partial^2 V_{eff}}{\partial \sigma \partial \sigma^\star } = \frac{1}{\lambda_r[\mu^2]} -\frac{1}{ 8 \nu {\pi} } \ln{\frac{m^4}{\mu^4}} . \label{l2d} 
\eq
In two dimensions, since $H[2]$ = $\frac{1}{2 \pi}$, we obtain the same RG equation for $\lambda$ as found in \cite{Zorzano},  namely

\bq
 \mu  \frac { d \lambda_r}{d \mu} = \frac{\lambda_r^2}{ 2 \pi \nu} .
 \eq
This is equivalent to the result from Eq. \eqref{l2d} that
\bq
\frac{1}{\lambda_r[\mu^2]}-\frac{1}{\lambda_r[{\mu'} ^2]} =  \frac{1}{ 8 \nu {\pi} } \ln{\frac{{\mu'}^4}{\mu^4}}.
\eq

\subsection{d=1}
At d=1, the first derivatives of the effective potential  are finite and we can just evaluate them and integrate once to obtain the effective potential.
We have
\ba
\frac {\partial V_{eff}}{\partial \sigma} && =  
  (\phi^\star)^2 -  \frac{1}{\lambda}  \sigma^\star  \nonumber \\
   +&&( 1-\sigma^\star) \frac{4}{\nu} \frac{ \Gamma \left(\frac{5}{4}\right)^2}{\pi ^{3/2}}\left(\frac{4  \sigma(1-\sigma^\star)} {\nu^2} \right)^{-1/4} \nonumber \\ 
\ea
\bq
\frac {\partial V_{eff}}{\partial \sigma^\star}  =  \phi^2 -  \frac{1}{\lambda}  \sigma -  \sigma   \frac{4} {\nu} \frac{ \Gamma \left(\frac{5}{4}\right)^2}{\pi ^{3/2}}\left(\frac{4  \sigma(1-\sigma^\star)} {\nu^2} \right)^{-1/4} .
\eq

We obtain:
\ba
V_{eff}&& = 
 (\sigma^\star - 1) \phi^2 +   \sigma (\phi^\star)^2     -  \frac{1}{\lambda} \sigma^\star \sigma \nonumber \\
 && +  \frac{ 4 \nu }{3}\frac{\Gamma \left(\frac{5}{4}\right)^2}{\pi ^{3/2}} \left(\frac{4 \sigma(1-\sigma^\star)} {\nu^2} \right)^{3/4} . 
\ea

For the renormalized coupling constant we then get
\bq
\frac{1}{\lambda_r[m^2]} = - \frac {\partial^2 V_{eff}}{\partial \sigma \partial \sigma^\star } = \frac{1}{\lambda} +  \frac{ 16 }{3  \nu} \frac{\Gamma \left(\frac{5}{4}\right)^2}{\pi ^{3/2}} \left(\frac{4  \sigma(1-\sigma^\star)} {\nu^2} \right)^{-1/4} .
\eq

For  the renormalized coupling at scale $\mu$ 
\bq
 \frac{1}{\lambda_r[\mu] }= \frac{1}{\lambda} +  \frac{1  }{\nu  \mu } H[1], 
\eq
where
\bq
H[1] =  \frac{ 16   }{3 } \frac{\Gamma \left(\frac{5}{4}\right)^2}{\pi ^{3/2}}. 
\eq 
Again forming the dimensionless  reaction rate $g_0 = \lambda/(\mu \nu) $,  the dimensionless renormalized reaction rate is 
\bq
g_r [\mu]= \frac{g_0[\mu]}{1+   g_0[\mu] H[1]},
\eq
 leading to the $\beta$ function:
\bq
\beta_g = \mu \frac{\partial g_r}{\partial \mu} = - g_r + g_r^2  H[1] .
\eq
For $d<2$ there is an infrared fixed point
\bq
g^\star = 1/H[1] .
\eq
To obtain the effective potential as a function of $\phi$ and $\phi^\star$ we need to solve for $\sigma$ and $\sigma^\star$ using the gap equations:
\bq
\frac {\partial V_{eff}}{\partial \sigma}  = \frac {\partial V_{eff}}{\partial \sigma^\star} =0.
\eq
To obtain the effective potential only as a function of $\phi$ we would need to solve for the constraint field $\phi^\star$ in terms of $\phi$ after eliminating $\sigma$ and $\sigma^\star$ using the gap equations. 

\section{ connection with KPZ equation}

The KPZ  \cite{kpz} equation is described by the stochastic partial differential equation:
\bq
\partial_t \phi(x,t) - \nu \nabla^2 \phi(x,t) =  \frac{\lambda}{2} (\nabla \phi )^2 + \eta(x,t). 
\eq
For simplicity we will assume Gaussian noise
\bq
\langle \eta (x,t) \eta (x',t') \rangle = \calA \delta (t-t') \delta^d(x-x') .
\eq

  For the KPZ equation, the physical ``gauge invariant" degree of freedom is $\nabla \phi$  which plays the role of 
an ``Electric Field".  For the KPZ equation
\bq
F[\phi] =  \frac {\lambda^2 }{2}  (\nabla \phi)^2.
\eq

For noise sources whose correlation functions  are translation invariant and local in time the KPZ equation has the Galilean symmetry:
\ba
&& {\vec x'} ={\vec x} +  \lambda  {\vec v} t, \nonumber \\
&&t = t', \nonumber \\
&&  \phi'(x,t)= \phi(x,t) - {\vec v} \cdot {\vec x} + \frac{\lambda}{2} v ^2 t.
\ea

As discussed in \cite{Lassig, Cardy} one can convert the KPZ  Langevin equation using a Cole-Hopt transformation to a Langevin equation  very similar to the one we have just discussed for the annihilation process $A+A \rightarrow 0$. 
We start by making a Cole-Hopf transformation:
\bq
\phi \rightarrow   \frac{2 \nu}{\lambda} \ln w.
\eq

Then $w$ satisfies the linear equation with multiplicative noise:
\bq
\partial_t {w}- \nu \nabla^2 w = \frac{\lambda} {2 \nu} w \eta.
\eq
This  stochastic differential equation can be obtained  from the  MSR Lagrangian which (apart from the Jacobean) is related to the functional delta function in changing variables from $\eta$
to $w, w^\star$.  
\bq
S_0 = \int dx dt \left[ w^\star \left(\partial_t {w}- \nu \nabla^2 w -\frac{\lambda} {2 \nu} w \eta \right) \right] .
\eq
Following the methodology discussed in Section II, we can now perform the integration over $\eta$ to obtain
\bq
Z[j, j^\star] = \int \calD w \calD w^\star \exp \left[- S_1 [w, w^\star] + w^\star j + w j^\star \right] ,
\eq
where 
\bq
S_1=  \int dx dt \left[ w^\star \left(\partial_t {w}- \nu \nabla^2 w \right) - \frac{\lambda^2 \calA}{2 \nu^2} (w^\star)^2 w^2  \right] .\label{action1a}
\eq
To compare with the annihilation process $A+A \rightarrow 0$ we introduce a new coupling constant:
\bq
  {\tilde \lambda} = \frac{\lambda^2 \calA}{2 \nu^2} .
  \eq
We can introduce the composite fields $\sigma$ and $\sigma^\star$ via a Hubbard-Stratonovich transformation.  That is we add to the above Lagrangian the action
\bq
S_{HS}  =\int dx dt \left[ \frac{1}   {\tilde \lambda} (\sigma^\star -   {\tilde \lambda} S^\star S^ \star)(\sigma -   {\tilde \lambda} S S ) \right].
\eq
Adding this to Eq.  \eqref{action1a}  we get an  action which again quadratic in $S,  S^\star$.
\ba
S_{2}[S, S^\star ,\sigma, \sigma^\star] && = \int dx \int_0^\tau dt \left[ S^\star \partial_t S+  \nu \nabla S^\star \nabla S  \right. \nonumber \\
&& \left.  -     (\sigma^\star ) S^2 -     \sigma (S^\star)^2     +   \frac{1} {\tilde \lambda} \sigma^\star \sigma \right] .
 \label{action2a}
\ea
We see apart from a sign change in the coupling constant  and the absence of a term linear in $S^2$  the Cole-Hopf form of the KPZ  actions is quite similar to that for $A+A \rightarrow  0.$ 

 Again introducing the renormalized coupling constant at scale $\mu$ , $  \tilde \lambda_r[\mu] $
 \bq
\tilde \lambda_r[\mu] = Z_g[\mu] \tilde \lambda; ~ Z_g^{-1} =  1 -\frac{ \tilde \lambda}{\nu} \Sigma[\mu^2],  \label{scat2}
 \eq
where
 \bq
 \Sigma[m^2 ]  = \Sigma_1[m^2]- \Sigma_2 [m^2] =   m^{d-2} \frac{1}{2-d}  H[d] ,
 \eq
and
 \bq
H[d] =\frac{2^{-d-1} (d+2) \pi ^{-\frac{d}{2}-\frac{1}{2}} \Gamma
   \left(\frac{3}{2}-\frac{d}{4}\right) \Gamma \left(\frac{d}{4}\right)}{\Gamma
   \left(\frac{d}{2}\right)}.
    \eq
From Eq. \eqref{scat2} we obtain the renormalization group equation : 
 \bq
 \mu  \frac { d \tilde \lambda_r}{d \mu} = - \frac{ \tilde \lambda_r^2}{\nu} H[d] .
 \eq

Again introducing  the dimensionless renormalized  coupling constant 
\bq
g_r =  \frac{\tilde \lambda_r} {\nu} \mu^{d-2},
\eq
the equation for $\beta_g$ is 
\bq
\beta_g = \mu \frac{\partial g_r}{\partial \mu} = (d-2) g_r - g_r^2 H[d] .
\eq
For $d>2 $ there is an unstable UV fixed point
\bq
g^\star = \frac{d-2} {H[d]} .
\eq
This leads to the roughening transition at $d=2$ as discussed in \cite{Cardy}. Note that our answer for $g^\star$  differs from the exact answer by the ratio $B[d]/H[d]$ which is plotted in Fig \ref{fig:fig1}. 

\section{Ginzburg-Landau model} 
The Ginzburg-Landau model  is the prototypic relaxation model of an Ising ferromagnet.   To make contact with Cardy's lecture notes \cite{Cardy} , we work in {\it reduced units}, so that $kT_c=1$ and use his notation that $\nu = D$.   This model is described by an equilibrium Hamiltonian

\bq
{\cal H} = \int d^d x \left[ \frac{1}{2} (\nabla S)^2 + \frac{1}{2} r_0 S^2 + \frac{1}{4} u S^4 \right] ,
\eq
where $r_0 \propto (T-T_{MF} )$.
The Langevin equation which relaxes to the equilibrium distribution determined from ${\cal H}$  is 
\ba
\partial_t S(x,t)&& = - D \frac{\delta {\cal H} }{\delta S(x,t)} + \eta(x,t). \nonumber \\
&& = D \left( \nabla^2 S- r_0 S -u S^3 \right) .
\ea
To satisfy the Einstein relation, one requires
\bq
\langle \eta(x,t)  \eta(x',t') \rangle = 2 D \delta^d (x-x') \delta (t-t').
\eq
In line with our discussion of the KPZ equation generating functional we now have
\bq
S_0 = \int dx dt \left[ S^\star \left(\partial_t {S}- D \nabla^2 S +r S + \lambda S^3 - \eta \right) \right],
\eq
where $r = D r_0$, $\lambda = D u$.
Following the methodology discussed in Section II, we can now perform the integration over $\eta$ to obtain
\bq
Z[j, j^\star] = \int \calD S \calD S^\star \exp \left[- S_1 [S, S^\star] + S^\star j + S j^\star \right] ,
\eq
where 
\bq
S_1=  \int dx dt \left[ S^\star\left(\partial_t {S}- D \nabla^2 S + r S + \lambda S^3   \right) - D S^\star S^\star  \right] .\label{action2b}
\eq
We can introduce the composite field $\sigma$ and Lagrange multiplier field  $\sigma^\star$  into the theory by the use of the functional delta function and the identity operator. That is we have 
\ba
1&&= \int \calD \sigma \delta ( \sigma - \lambda S^2 -r)  \nonumber \\
&&= \int \calD  \sigma^\star \calD \sigma  \exp [ \frac {1}{\lambda} \sigma^\star (\sigma- \lambda S^2 - r )].
\ea

Adding this to Eq.  \eqref{action2b}  we get an  action which again quadratic in $S , S^\star$.
\ba
&&S_{2}[S,S^\star,\sigma]  = \int dx \int_0^\tau dt  \left[ S^\star\left(\partial_t {S}- D \nabla^2 S  \right.  \right. \nonumber \\
 && \left. \left.  + \sigma S  \right) - D S^\star S^\star +\sigma^\star S^2 -\frac{ \sigma^\star }{\lambda}(\sigma- r)\right] .
 \label{action3a} \nonumber \\
\ea
The ``bare''  matrix inverse propagator can be written 
\ba
&&G^{-1}(x-x',t-t'; \sigma, \sigma^\star) =   \nonumber \\
&& \left(  \begin{array}{cc}
 2 \sigma^\star    & - \partial_t - D \nabla^2 + \sigma\\ 
  \partial_t - D \nabla^2 + \sigma    & - 2 D    \\ 
  \end{array}
\right)  \delta^d (x-x') \delta (t-t') . \nonumber \\
\ea

Performing the integration over the fields $S, S^\star$ and adding sources for the auxiliary fields we obtain
\ba
&&Z[j,j^\star, J, J^\star]= \nonumber \\
&& \int \calD \sigma \calD \sigma^\star \exp \left(-  \tilde S [ \sigma,\sigma^\star, j, j^\star] 
+J^\star \sigma +\sigma^\star J \right), \nonumber \\
\ea
where schematically: 
\bq
\tilde S = - \int \left(j \cdot G \cdot j  -\frac{ \sigma^\star }{\lambda} (\sigma- r) \right)  + \frac{1}{2}  \rm {Tr} \ln  G^{-1}(\sigma,\sigma^\star) .
\eq
 Performing the integration over the auxiliary fields by steepest descent  and keeping the saddle point contribution, we can then Legendre transform the result for $\ln Z$. We then obtain in the LOAF approximation
\bq
\Gamma [ S, S^\star, \sigma, \sigma^*] = S_2[S, \sigma] + \frac{1}{2}  \rm {Tr} \ln  G^{-1}(\sigma,\sigma^\star) .
\eq
Here $\Gamma$ is the generating functional of the one particle irreducible graphs.
Restricting ourselves to constant fields and defining the effective potential as 
\bq
V_{eff} = \frac{ \Gamma [S,S^\star, \sigma, \sigma^\star]} {\Omega},
\eq
where $\Omega$ is the space time volume,  we obtain 
\ba
V_{eff} && =   \sigma S^\star S - D S^\star S^\star + \sigma^\star S^2 - \frac{\sigma^\star}{\lambda} (\sigma-r)  \nonumber \\
 && +\frac{1}{2}  \rm {Tr} \ln  G^{-1}(\sigma,\sigma^\star) .
\ea
This consists of  the classical potential written in terms of the auxiliary fields  plus the saddle point contribution to the fluctuations. For static fields we can Fourier transform the propagator and evaluate the fluctuation contribution by calculating the log of the determinant of $G^{-1}$ in momentum space.
In momentum space we have 
\ba
&&G^{-1}(\omega, q, \sigma, \sigma^\star) =  \nonumber \\
&& \left(  \begin{array}{cc}
   2 \sigma^\star   &~~ - i \omega + D k^2 + \sigma\\ 
i \omega + D k^2 + \sigma     &~~ - 2 D    \\ 
  \end{array}
\right) .
\ea

Evaluating the determinant we obtain
\ba
&&V_{eff}  =   \sigma S^\star S - D S^\star S^\star + \sigma^\star S^2 - \frac{\sigma^\star}{\lambda} (\sigma-r)  \nonumber \\
&& + \frac{1}{2} \int \frac{d^d k}{(2 \pi)^d} \int_{-\infty}^{\infty} \frac{d \omega}{2 \pi} \ln\left[\omega^2 +(Dk^2+\sigma)^2 + 4 D \sigma^\star)  \right] . \nonumber \\
\ea
Performing the integration over $\omega$ we obtain
\ba
&&V_{eff}  =  \sigma S^\star S - D S^\star S^\star + \sigma^\star S^2 - \frac{\sigma^\star}{\lambda} (\sigma-r)  \nonumber \\
&& + \frac{1}{2} \int \frac{d^d k}{(2 \pi)^d} \sqrt{(Dk^2+\sigma)^2 + 4 D \sigma^\star} .
\ea
This expression needs to be evaluated at the solution of the gap equations 
\bq
\frac {\partial V_{eff}}{\partial \sigma} = \frac {\partial V_{eff}}{\partial \sigma^\star} = 0.
\eq

We have
\ba
\frac {\partial V_{eff}}{\partial \sigma^\star} && =  S^2- \frac{(\sigma-r)} {\lambda}  \nonumber \\
&&+  D  \int \frac{d^d k}{(2 \pi)^d} \left((Dk^2+\sigma)^2 + 4 D \sigma^\star  \right)^{-1/2}.
\ea
\ba
&& \frac {\partial V_{eff}}{\partial \sigma}  =  (S^\star)^2 - \frac{\sigma^\star} {\lambda}  \nonumber \\
&&+  \frac{1}{2}   \int \frac{d^d k}{(2 \pi)^d} \left(Dk^2+ \sigma \right) \left((Dk^2+\sigma)^2 + 4 D \sigma^\star  \right)^{-1/2}. \nonumber \\
\ea

The equation for the renormalized coupling constant is 
\ba
&&\frac{\partial^2 V_{eff}} {\partial \sigma \partial \sigma^\star} = -\frac{1}{\lambda_r[m_1^2,m_2^2] }  \nonumber \\
&& =- \frac{1}{\lambda} -\frac{1}{D} \frac{\Omega_d}{(2 \pi)^d}   \int dk  k^{d-1} \frac{k^2+ m_1^2}{ \left[ (k^2+ m_1^2)^2 + m_2^4 \right]^{3/2}}, \label{v2gl} \nonumber \\
\ea
where  $m_2^4 =  4 \sigma^\star/D ,~~ m_1^2 = \sigma/D $. 
We see the critical dimension is $d=4$. 
If we expand the integral in Eq. \eqref{v2gl} as a power series in $m_2^4$ around zero, we notice only the first term ($m_2^4=0$) is divergent for $d=4$. 
 This suggest that we can define a renormalized  coupling constant at $m_2^4 =0$ and $m_1^2 = \mu^2$  via
\ba
- \frac{1}{\lambda[\mu^2] }&&=- \frac{1}{\lambda} -\frac{1}{D} \frac{\Omega_d}{(2 \pi)^d}   \int dk  k^{d-1} \frac{1}{(k^2+ \mu^2)^2} \nonumber \\
&& = - \frac{1}{\lambda} -\frac{1}{D} \frac{\Omega_d}{2 (2 \pi)^d}  (\mu)^{(d-4)}  \Gamma(d/2) \Gamma(2-d/2). \nonumber \\
\ea
We can expose the pole in the Gamma function and rewrite this equation as
\bq
\lambda_r[\mu^2] = \frac{\lambda}{1+ \frac { \lambda \mu^{d-4}}{D(4-d)}  J[d]},
\eq
where 
\bq
J[d] = \frac{\Omega_d}{(2 \pi)^d} \Gamma[d/2] \Gamma[3-d/2] .
\eq
Introducing the dimensionless bare coupling constant
\bq
g_0[\mu^2] = \lambda \mu^{d-4} /D,
\eq
then the dimensionless renormalized coupling constant is
\bq
g_r[\mu^2] = \frac{g_0[\mu^2]} {1+  \frac{g_0[\mu^2]}{4-d} J[d]}.
\eq
The $\beta$ function for $g_r[\mu^2] $ is given by
\bq
\beta[g_r] = \mu  \frac{\partial g_r[\mu^2]}{\partial \mu} = [d-4] g_r[\mu] + g_r[\mu^2]^2 J[d] .
\eq
There is a fixed point  for $d \leq 4$ at
\bq
g^\star =  \frac{\epsilon}{J[d]},
\eq
where $\epsilon = 4-d$.  This is quite similar to the result of  a perturbative analysis of the problem with the distinction that $J[d]$ is replaced by a related $d$ dependent
function $K[d]$ \cite{Cardy}. 

As in the previous examples we can now regulate the second derivative of $V_{eff}$ using the definition of $\lambda[\mu]$, so that
\ba
&&\frac{\partial^2 V_{eff}} {\partial \sigma \partial \sigma^\star} =- \frac{1}{\lambda[\mu] } -\frac{1}{D} \frac{\Omega_d}{(2 \pi)^d}    \times \nonumber \\
&&  \int dk  k^{d-1} \left[ \frac{k^2+ m_1^2}{ \left[ (k^2+ m_1^2)^2 + m_2^4 \right]^{3/2}}-  \frac{1}{(k^2+ \mu^2)^2} \right]. \nonumber \\
\ea

This can be evaluated analytically as a power series in $m_2^4$ for example and then one can reconstruct the full effective potential by integrating this result with respect to $\sigma$ and $\sigma^\star$ and adding the appropriate classical terms.  Evaluating the resulting  effective potential at the solution of the gap equation then gives the effective potential in terms of $S, S^\star$.  Finally, solving for the Lagrange multiplier field $S^\star [S]$ and substituting that in the potential gives the Onsager-Machlup form of the potential discussed in our previous paper.

 \section{Conclusions}
 In this paper we discussed two ways that the MSR formalism arises in stochastic systems.  One way, is when the system is describable in terms of birth-death processes via a Master equation.  The second is when there is specified external noise applied to a classical evolution equation.  For both cases we show how to obtain the effective action and effective potential as a loop expansion in an auxiliary composite field.  This results in lowest order in a self consistent approximation which allows one to explore coupling constant regimes not accessible in the usual text-book semi-classical loop expansion.  We considered three problems:  the classical annihilation process $A+ A \rightarrow 0$, the Kardar-Parisi-Zhang equation in Cole Hopf transformed form, and the Ginzburg-Landau equation for relaxation to equilibrium dynamics. In each case we determined the running of the coupling constant as a function of the dimension $d$  and calculated the $\beta$ function which we compared to exact and  perturbation theory results.  We obtained quantitatively accurate results for the fixed points of both the annihilation process and the KPZ equation in the Cole-Hopf transformed form.  More importantly we indicated a way to calculate in a well prescribed manner (as a loop expansion in the auxiliary field propagators)  the corrections to the lowest order (LOAF) approximation.  This formalism is well suited for determining the time evolution of the noise averaged correlation functions.  This type of approximation was very fruitful in discussing the dynamics of phase transitions as exemplified by the studies of both chiral phase transitions in particle physics \cite{chiral} and phase separation transitions in BECs \cite{Chien}.

\begin{acknowledgments}
I would like to thank Juan Perez-Mercader for thoughtful discussions and suggesting this research. I  would also like to thank Gourab Ghoshal, John Dawson and  Jean-Sebastien Gagnon for  valuable discussions. 

\end{acknowledgments}

\end{document}